\documentstyle[12pt,a4]{article}

\font \msb=msbm10 scaled \magstep1

\newcommand{\bC}{\mbox{\msb C} }

\font \eul=eufm10 scaled \magstep2

\newcommand{\ggl }{\mbox{\eul g}}

\def\a{{\alpha}}
\def\D{{\Delta}}
\def\te#1{{\tilde{#1}}}
\def\g{{\gamma}}
\def\s{{\sigma}}

\def\nn{ \nonumber }
\def\bq{ \begin{equation} }
\def\eq{ \end{equation} }
\def\ben{ \begin{eqnarray} }
\def\en{ \end{eqnarray} }
\def\ll{ \label }
\def\dfrac#1#2{{\displaystyle{#1\over#2}}}

\newtheorem{prop}{Proposition}
\begin{document}
\title{
\begin{flushleft}
{\small{\sl Submitted to:} {\sf Physics Letters A}}\\
\end{flushleft}\vskip 3cm
The Kowalewski top in the SUSY quantum mechanics.}
\author{Tsiganov, A. V.\\
{\small\it
  Department of Mathematical and Computational Physics,
 Institute of Physics,}\\
{\small\it
 St.Petersburg University,
 198 904,  St.Petersburg,  Russia}\\
{\small\it e-mail: atsiganv@snoopy.niif.spb.su}}
\date{}
\maketitle

\begin{abstract}
The Kowalewski top on Lie algebras $o(4)$, $e(3)$ and $o(3,1)$
is embedded in the SUSY quantum mechanics. In two dimensions we
give the new prescription for construction of the pairs of
integrable systems by using a standard SUSY algebra. At the
proposed scheme the Goryachev-Chapligin top is shown to be a
natural partner of the Kowalewski top.
\end{abstract}
\vskip1cm
Supersymmetrical quantum mechanics represents a concise algebraic
form of the spectral equivalence between different Hamiltonian
quantum systems realized by means of the Darboux transformation
\cite{aj95}.  While a theory of the one-dimensional supersymmetrical
quantum mechanics is now well developed, this is not so for the
multidimensional ones and we still are in phase of case studies.  The
aim of this note is to show that the Kowalewski top and the
Goryachev-Chapligin top are the two-dimensional supersymmetric models
in quantum mechanics. Notice, that the construction of
isospectral two-dimensional Hamiltonians in supersymmetrical
quantum mechanics is tightly connected with another problem,
namely with a search for the second integral of motion for the
quantum integrable systems. Moreover, recently the SUSY algebra
have been applied to the separation of variables for the
two-dimensional hamiltonian systems \cite{aj95,rwt95}.

We recall here the basic algebraic constructions of the
supersymmetrical quantum mechanics. All details can be found in
\cite{aj95} and references therein.  The interwining relations
between two Hamiltonians $h^{(1)}$ and $h^{(2)}$ with equivalent spectra
are given by
\ben
&h^{(1)}q^+=q^+h^{(2)}\,,\qquad
&q^-h^{(1)}=h^{(2)}q^-\,,\nn\\
&h^{(1)}\Psi_n^{(1)}=E_n\Psi_n^{(1)}\,,\qquad
&h^{(2)}\Psi_n^{(2)}=E_n\Psi_n^{(2)}\,,\ll{ss1}\\
&\left(q^+\right)^{\dag}=q^-\,,&\nn
\en
The concise algebraic form of the spectral equivalence is given
by the superalgebra for the partners $h^{(1)}$ and $h^{(2)}$
and off-diagonal supercharges
\ben
&H=\left(\begin{array}{cc}h^{(1)}&0\\0&h^{(2)}
\end{array}\right)\,,\qquad
&Q^+=\left(Q^-\right)^{\dag}=
\left(\begin{array}{cc}0&0\\q^-&0
\end{array}\right)\,,\nn\\
\ll{ss2}\\
&\left(Q^+\right)^2=\left(Q^-\right)^2=0\,,
\qquad &\left[H,Q^\pm\right]=0\,.\nn
\en
Since the super charges $Q^\pm$ commute with the Hamiltonian
$H$  one expects that the closing of the superalgebra will
lead to the symmetry operator $R$ (central charge) \cite{aj95}
\bq
\{Q^+,Q^-\}=R\,,\qquad [H,R]=0\,,\qquad
R=\left(\begin{array}{cc}R_1&0\\0&R_2
\end{array}\right)\,,\ll{cc}
\eq
where braces $\{, \}$ mean anticommutator of quantum operators.  In
the two-dimensional case the closing of the SUSY algebra leads to
the integrability of the corresponding dynamical system and $R$ is
the second integral of motion.

Let us consider quantum (or classical)
dynamical systems on algebra $\ggl=o(4)$, $e(3)$, $o(3,1)$
with generators obeying the following commutator relations:
 \ben
 &&\bigl[ l_i\,, l_j\,\bigr]= -i\eta\varepsilon_{ijk}\,l_k\,,
 \qquad
 \bigl[ l_i\,, g_j\,\bigr]= -i\eta\varepsilon_{ijk}\,g_k\,,
 \nn\\
\ll{gen}\\
 &&\bigl[ g_i\,, g_j\,\bigr]= i\eta P\varepsilon_{ijk}\,l_k\,, \qquad
 i, j=1, 2, 3.
 \nn
 \en
here $\eta$ is a Plank constant and the constant $P$ is given by
\ben
P&=-1\quad &\ggl=o(4)\,,\nn\\
&=0\quad &\ggl=e(3)\,,\ll{p}\\
&=1\quad &\ggl=o(3,1)\,,\nn
\en
The Casimir operators
\bq
a^2=g_kg_k-Pl_kl_k\,,\qquad
l=g_kl_k \,,\ll{cas}
\eq
are supposed to be fixed.
The Hamiltonian for the Kowalewski top is
\bq
h=l_1^2+ l_2^2+2 l_3^2-i(\a_1  g_++\a_2 g_-)
\,,\qquad\a_k\in\bC\,.
\ll{hk}
\eq
Here we use the general form for the Hamiltonian introduced in
classical mechanics \cite{hh87,brs89}.  Following Kowalewski we
introduce new variables
\ben &&q^-=l_-^2+2i\a_1g_--\a_1^2 P\,,\nn\\
\ll{qpm}\\ &&q^+=l_+^2+2i\a_2g_+-\a_2^2 P\,,\nn
\en
with $l_\pm=l_1\pm il_2$ and $g_\pm=g_1\pm ig_2$. In classical mechanics
these variables have been introduced in algebro-geometric approach
for linearization of the Kowalewski flow \cite{hh87}.

\begin{prop}
In quantum mechanics operators $q^{\pm}$ are supercharges of
SUSY algebra (\ref{ss1}) for the following isospectral
Hamiltonians
\bq
h^{(1)}=h+2\eta l_3\,,\qquad h^{(2)}=h-2\eta l_3\,,\ll{ss12}\\
\eq
where $h$ is the Hamiltonian of the Kowalewski top (\ref{hk}).
\end{prop}
It can be proved by using the commutator relations introduced in
\cite{kom87}.

For the Kowalewski top one finds that the central charge $R$ is
a diagonal operator with the following components
\bq
R_1=q^+q^-\,,\qquad R_2=q^-q^+\,,\qquad
[R_j,h^{(j)}]=0\,,\ll{comp}
\eq
which are the symmetry operators of the Hamiltonians $h^{(1)}$
and $h^{(2)}$, respectively.  From the SUSY algebra (\ref{ss2})
we get the pair of commuting operators
\bq
h=\dfrac12(\,h^{(1)}+h^{(2)}\,)\,,\qquad
k=\dfrac12\{q_+, q_-\}+2\eta^2\{l_+, l_-\}\,,
\ll{k}
\eq
which are Hamiltonian and second integral of motion
for the Kowalewski top.

Now we shall construct a new two-dimensional integrable system
by using the known SUSY algebra for the Kowalewski top.
The SUSY algebra (\ref{ss1}) can be rewritten as
\bq
[q^\pm,(\,h^{(1)}+h^{(2)}\,)\,]=\pm\{\,(h^{(1)}-h^{(2)}\,),q^\pm\,\}
\ll{hq1}
\eq
or, by using the concise notations, as
\bq
[q^\pm,h]=\pm\{\,f,q^\pm\,\}\,,
\qquad h=h^{(1)}+h^{(2)}\,,\quad
f=h^{(1)}-h^{(2)}\,.
\ll{hq2}
\eq
\begin{prop}
Let we have the SUSY algebra (\ref{ss2}) for certain
two-dimensional integrable system defined by four operators
$h,f$ and $q^{\pm}$.  If the following equation in $\D h$ can
be solved
\bq
[\D h,[q^+,q^-]\,]=[f,\{q^+,q^-\}\,]\,,\ll{eq}
\eq
than the pair of the mutually commuting operators
\ben
\te{h}&=&h+\D h=h^{(1)}+h^{(2)}+\D h\,,\nn\\
\ll{ns}\\
\te{k}&=&[q^+,q^-]\,,\qquad[\te{h},\te{k}]=0\,,\nn
\en
defines a new two-dimensional  integrable system.
\end{prop}
It can be simply proved by using the Jacobi indentity for
commutator relations.
The new equation (\ref{eq}) can be rewritten as
\ben
[\D h,[q^+,q^-]\,]&=&\dfrac12[f, R_1+R_2]
=\dfrac12\,{\rm tr}\,\left[\,(\s_3 H)\otimes R- R\otimes(\s_3H)\right]\nn\\
\nn\\
&=&-\dfrac12\,{\rm tr}\,\left[\,H\otimes (\s_2 R)- (\s_3R)\otimes H\right]\nn\\
\nn\\
\s_3&=&\left(\begin{array}{cc}1&0\\0&-1
\end{array}\right)\,,\nn
\en
where we used a tensor product of $2\times 2$ matrices $H,\,R$
(\ref{ss2}-\ref{cc}) and $\s_3$ is a Pauli matrix.

The Kowalewski top with the integrals of motion $(h,k)$
(\ref{k}) and with the SUSY algebra defined by (\ref{ss1}) has
the integrable partner with integrals of motion
$(\te{h},\te{k})$ given by (\ref{ns}) in the one-parameter
subset of orbits $\cal O$  ($a^2=const$, $l=l_kg_k=0$
(\ref{cas})) of the Lie algebra $\ggl=e(3)$. It is the
Goryachev-Chapligin top with the following integrals of motion
\ben
\te{h}&=&l_1^2+ l_2^2+4 l_3^2-2i(\a_1  g_++\a_2 g_-)\,,\ll{gc}\\
\nn\\
\te{k}&=&-8i\left(l_3(l_1^2+l_2^2+1/4) +i\{g_3,(\a_1 l_++\a_2 l_-)\}\right)\,,\nn\\
\nn\\
{\rm where}\qquad \D h&=& 2l_3^2-i(\a_1  g_++\a_2 g_-)\,,\nn
\en
which is the partial solution of (\ref{eq}) on the
corresponding orbit of $e(3)$. The another relations of these
systems in framework of inverse scattering method are discussed
in \cite{bk88,ts95}.

Another example of the application of the Proposition 2 to
the two-dimensional systems can be found in \cite{rwt95}.

By definition of the Hamiltonians $h^{(j)}$ and of the
supercharges $q^\pm$ the generator $l_3$ is distinguished
generator and it can be changed to $l_3+\g$ without of
violation of the superalgebra (\ref{ss2}), here parameter $\g$
being arbitrary.  Therefore we can simple carry over all
results on the systems with the shifted generator $l_3$.  The
corresponding systems are the Kowalewski gyrostat and
Goryachev-Chapligin gyrostat \cite{brs89,kom87}.

The author would like to express his gratitude to S.
Rauch-Wojciechowski and to Department of Mathematics of
Linkoping University for warm hospitality.  This work was
supported by Swedish Royal Academy of Science NFR grant F-AA/MA
08677-326 in framework of a collaborative program and RFBR
grant 96-01-00537.


\end{document}